# Fourier-imaging of single self-assembled CdSe nanoplatelet chains and clusters reveals out-of-plane dipole contribution

*Jiawen Liu[1], Lilian Guillemeney[2], Arnaud Choux[1], Agnès Maître[1], Benjamin Abécassis[2] and Laurent Coolen[1]*

[1]Sorbonne Université, CNRS, Institut de NanoSciences de Paris, INSP, F-75005 Paris, France

[2]Univ Lyon, CNRS, École Normale Supérieure de Lyon, Laboratoire de Chimie UMR 5182 46 allée d'Italie, F-69007 Lyon, France

**Abstract :** Fluorescent semiconductor nanoplatelets (epitaxial quantum wells) can be synthesized with excellent monodispersity and self-assembled in highly-ordered structures. Modifications of their electronic and luminescence properties when stacked, due to strong mechanical, electronic or optical interactions between them, have been the topic of intense recent discussions. In this paper, we use Fourier imaging to measure the different dipole components of various nanoplatelet assemblies. By comparing different measurement conditions and corroborating them with polarimetric analysis, we confirm an excellent precision on the dipole components. For single nanoplatelets, only in-plane dipoles (parallel to the platelet plane) are evidenced. For clusters of 2-10 platelets and chains of 30-300 platelets, on the other hand, a clear out-of-plane dipole component is demonstrated. Its contribution becomes more significant as the number of platelets is increased. We review possible explanations and suggest that the added out-of-plane dipole can be induced by strain-induced nanoplatelet deformations.

Semiconductor nanoplatelets (NPLs)[1] are quasi-two-dimensional atomically flat nano-emitters with intense, spectrally monodisperse[2] and anisotropic[3-6] emission and potential applications in light-emitting devices[7], photovoltaic cells[8], field effect transistors[9] and lasers[10]. Fluorescence studies usually consider, as much as possible, isolated emitters, either by spectroscopy of dilute solutions or by single-molecule microscopy. Interactions between compact assemblies of stacked emitters are however receiving increased attention. In opto-electronics, for instance, these interactions must be understood as they can either be detrimental to a device's efficiency or allow new energy transfer strategies[8,11-12]. Semiconductor platelets are an ideal model system for studying such interactions because they have a pronounced tendency to spontaneous cofacial stacking[13-14] and because, when stacked, their interactions should be strong as they present high absorption cross-section[15-16], low Stokes shift and large in-plane dipoles[3,5-6] which can be aligned parallel at very short distance (2-5 nm) from each other[17-18]. For instance, a second emission peak appears at low temperatures for stacked NPLs and was tentatively attributed to phonon coupling[13], p-state emission[19], delocalized excimer states[20], surface states[21] and most recently to negatively-charged trions[22]. Förster resonance energy transfer (FRET) has also been evidenced between stacked platelets[8,23] with transfer rates[24] up to $(1.5\ ps)^{-1}$. By a careful control of the ligands and solvents, NPLs can be assembled into linear chains of a few hundred nm to a few µm, with good photostability and brightness, excellent order and highly uniform inter-NPL distance[25]. Ligand-induced strains within platelets can create a twist of these NPL chains[26-27] and may be responsible for the recently-observed permanent dipole parallel to the chain[28].

A key characteristic of a fluorescent system is the orientation of its radiating dipole, with implications for orientation and conformation tracking in bio-imaging[29] and super-resolution[30], for directional emission and light extraction in opto-electronics[12] or for coupling and Purcell efficiency in nanophotonics and plasmonics[31-32]. A corollary question is how many dipoles contribute to the emission. For instance, because their emission originates from two degenerate states of orthogonal dipole orientations, CdSe NPLs can be well described by an incoherent sum of two in-plane dipoles[3,5-6]. This can be an advantage, for instance, for light extraction in light-emitting diodes or for energy-funneling strategies in photovoltaics[3]. Other two-dimensional systems have also been investigated[47] in order to demonstrate vertical-dipole emission as it would provide enhanced coupling to plasmonic surfaces or photonic waveguides[33] or to Förster-transfer donors[34]. Vertical deposition of platelets by thick-shell growth[35] or by air-liquid interface self-assembly[4,34] was also demonstrated.

By combined polarimetric and Fourier-image analyses, the different dipoles components of a single fluorescent nano-emitter can be discriminated[6,35-36]. In this paper, we analyse the in-plane and out-of-plane dipoles of single isolated platelets and of NPL stacks of various lengths. We show that, while no out-of-plane contribution can be resolved (< 0.03) for isolated NPLs, a clear out-of-plane dipole appears and grows as NPLs tend to cluster and assemble into longer chains. We discuss various hypotheses in relation with the NPL's tilt and deformations and the chain's internal electric dipole.

**Results :**

Figure 1(a) shows a transmission electron microscopy (TEM) image of the CdSe nanoplatelets, synthesized as described in ref. 24. They present a well-controlled rectangular shape with 20-nm length and 7-nm width (with respective 4-nm and 2-nm dispersions). Some overlapping between platelets (darker grey shade) indicates a spontaneous tendency to stack together. Figure 1(c) plots the NPL absorption and emission spectra. The very low 2-nm Stokes shift is characteristic of NPLs. The 549-nm emission wavelength is associated with a 1.5-nm thickness : 6 Cd layers and 5 Se layers. Analysis of single-NPL emission spectra has shown that the inhomogeneous broadening is negligible for these samples[24]. Given their extreme shape anisotropy, the NPLs' emission wavelength is solely dependent on their thickness, which is controlled during the synthesis with monolayer precision[2].

Two different NPL chain samples were self-assembled as in ref. 24 (see Methods), one "long-chain" sample with length 1-2 µm (fig. 1(b)) and one "short-chain" sample with lengths 100-500 nm. The long chains present a twist which was discussed in ref. 26 and attributed to ligand-induced internal NPL strain. The NPL center-to-center distance can be estimated from high-resolution TEM images (see S.I. in ref. 24 for details) as 5.7 nm.

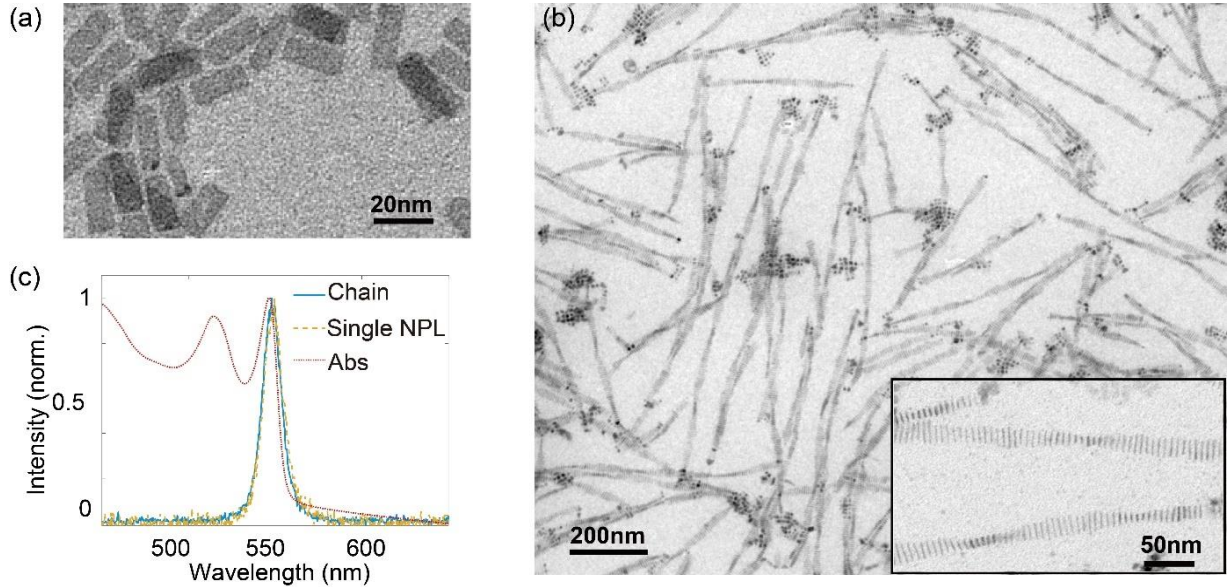

*Fig 1 : TEM images of (a) single and clustered platelets and (b) 1-2 µm NPL chains. (c) Absorption (cuvette measurement) and emission (microphotoluminescence) spectra of isolated NPLs, emission spectrum of NPL chains (microphotoluminescence).*

A CCD image of a single long chain under wide-field mercury-lamp illumination is shown on fig. 2(a). The sample consisted in a glass slide with emitters covered by a polymer (polymethylmetacrylate – PMMA), observed with a 1.4 numerical-aperture objective (fig. 2(b)). This configuration was chosen so that the radiation pattern was strongly dependent on the different emitting dipole contributions. For comparison, three different PMMA thicknesses were used : 130, 180 and 210 nm (measured with 5 nm precision by profilometry). We excited the chains at their centre and used Fourier-plane imaging (see Methods) to measure the radiation pattern of single NPL chains. Figure 2(c) shows the Fourier-plane images of three long chains under these different configurations. The radius corresponds to the maximal collected angle $\theta_{coll} = \sin^{-1}(N.A./n_{glass})$ = 69°. The maximal emission is directed along the chain axis, in agreement with the expectation that the in-plane NPL dipoles (which are perpendicular to the chain axis : direction Φ on fig. 2(c)) radiate mainly into their orthogonal direction.

The measured radiation patterns for typical chains are plotted in fig. 2(d) along planes parallel and orthogonal to the chain axis. These patterns are very different depending on the PMMA thickness because the collected electric field is a sum of direct emission and emission reflected by the glass-air interface. However, some common features arise : a lobe at normal 0° angle ; a sharp change or peak at the critical angle $\theta_c = \sin^{-1}(1/n_{glass})$ = 42° ; more or less strong lobes at $\theta > \theta_c$ corresponding to total internal reflection of the emission.

We fit the radiation patterns by an incoherent sum of three dipoles. For all samples considered in this paper, we will label ∥,1 the stronger NPL in-plane dipole, ∥,2 the weaker in-plane dipole, and ⊥ the out-of-plane dipole. Note that "in plane" will refer to the plane of the nanoplatelet. For a single NPL deposited horizontally on the substrate, in-plane dipoles ∥,1 and ∥,2 will be horizontal, while the out-of-plane direction ⊥ will be vertical. However for the NPL chain, dipole ⊥ is horizontal and parallel to the chain axis, while dipole ∥,1 is horizontal and orthogonal to the chain axis, and dipole ∥,2 is vertical (fig. 2(e)). We describe the weight of each contributing dipole by a coefficient $\eta$ proportional to the *square* of its dipole moment and normalized so that

$$\eta_{\parallel,1} + \eta_{\parallel,2} + \eta_{\perp} = 1$$

The fitted radiation patterns (calculated following ref. 37) are plotted as solid lines on fig. 2(d). The overall agreement between experimental and fitted curves is excellent. We estimate (Fig. S1) the general precision of the fitted values to ± 0.03 for the out-of-plane coefficient $\eta_{\perp}$ (which will be our main focus thereafter) and ± 0.05 for the weaker in-plane coefficient $\eta_{\parallel,2}$. Note that the angles $\theta \sim 0$ are subject to very strong noise because they are inferred from just a few pixels on the Fourier image. Some deviation from the model is observed at higher ($\theta > 60°$) angles and may be caused by objective aberrations. Similar $\eta$ coefficients are obtained for all chains. The radiation patterns with the different PMMA thicknesses, although qualitatively very different, can all be fitted with very close $\eta$ values : this is a striking confirmation of the quality of our measurement of the three dipole contributions.

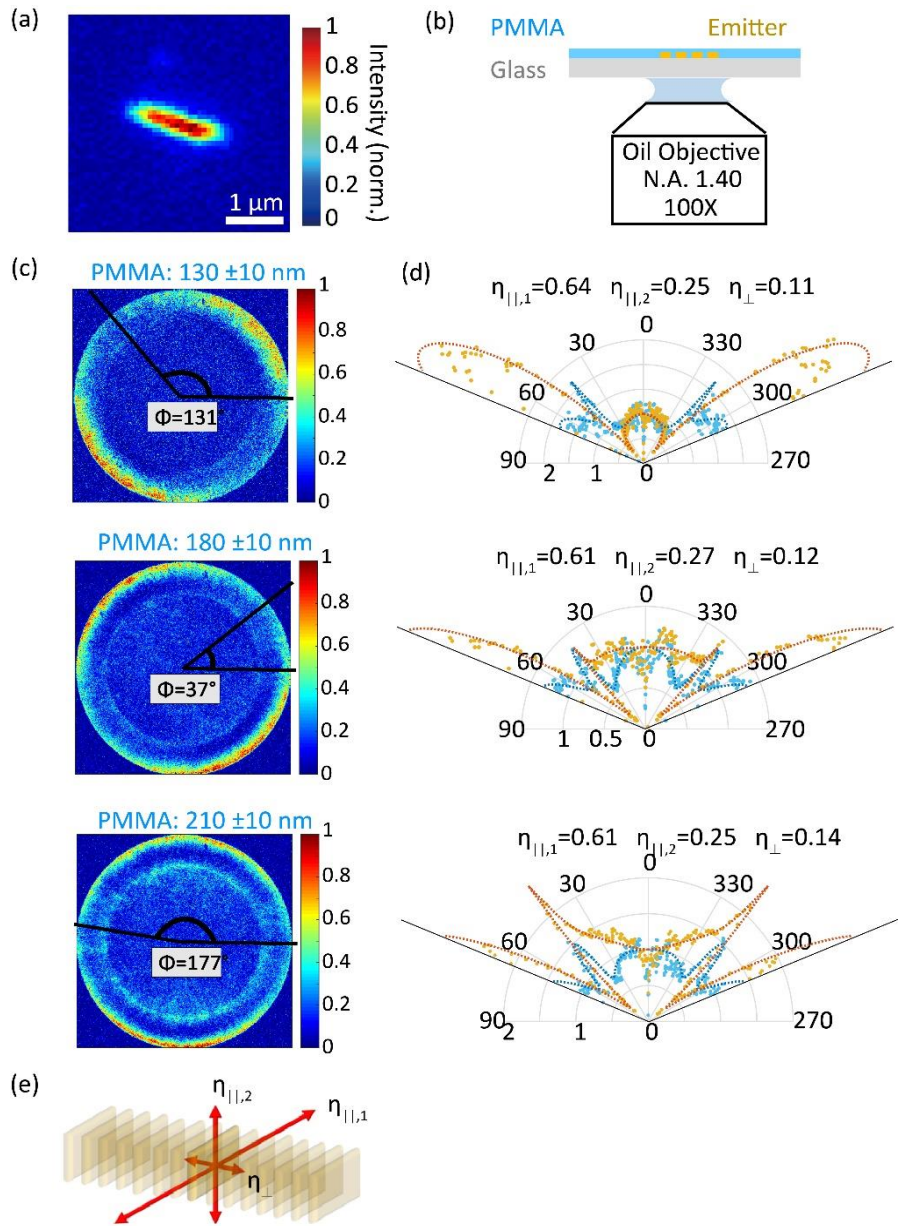

Figure 2 : (a) CCD image of a long NPL chain under wide-field mercury lamp illumination. (b) Schematic of the measurement configuration. (c) Obtained Fourier images of typical single NPL chains for three

*configurations with different PMMA thicknesses. The angle $\Phi$ indicates the orientation of the stronger dipole $\eta_{\parallel,1}$ in the chain obtained from polarization analysis. (d) Corresponding radiation patterns along axes perpendicular (orange curves, direction $\Phi$ on fig. 2(c)) and parallel (blue curves) to the chain axis. Dots : experimental data. Lines : fit by a sum of three dipoles. (e) Schematic of the three-dipole emission model.*

Further support for our measurement is provided by comparison with polarization analysis. Figure 3(a) shows a typical emission polarimetric curve (obtained as described in ref. 35) from which an experimental degree of polarization $\delta_{exp} = (I_{max} - I_{min})/(I_{max} + I_{min})$ = 0.30 can be fitted. Figure 3(b) shows the theoretical degree of polarization (calculation outlined in Methods) as a function of the dipole contributions $\eta_{\parallel,2}$ and $\eta_\perp$ (keeping in mind that $\eta_{\parallel,1} + \eta_{\parallel,2} + \eta_\perp = 1$). It can be noted that maximally-polarized emission is obtained when there is a single horizontal dipole ($\eta_{\parallel,1} = 1$ or $\eta_\perp = 1$) while the emission is unpolarised when there is a single vertical dipole ($\eta_{\parallel,2} = 1$) or when the two horizontal dipoles have the same moment ($\eta_{\parallel,1} = \eta_\perp$). For the parameters $\eta$ obtained from the Fourier image, the theoretical degree of polarization $\delta_{theo}$ = 0.30 is in perfect agreement with the measured degree of polarization $\delta_{exp}$ = 0.30. Figure 3(c) compares these values of $\delta$, experimental and theoretical (from $\eta$ factors), for all emitters considered in this paper. A remarkable agreement is always found between these two values, again validating the precision of our measured $\eta$ coefficients.

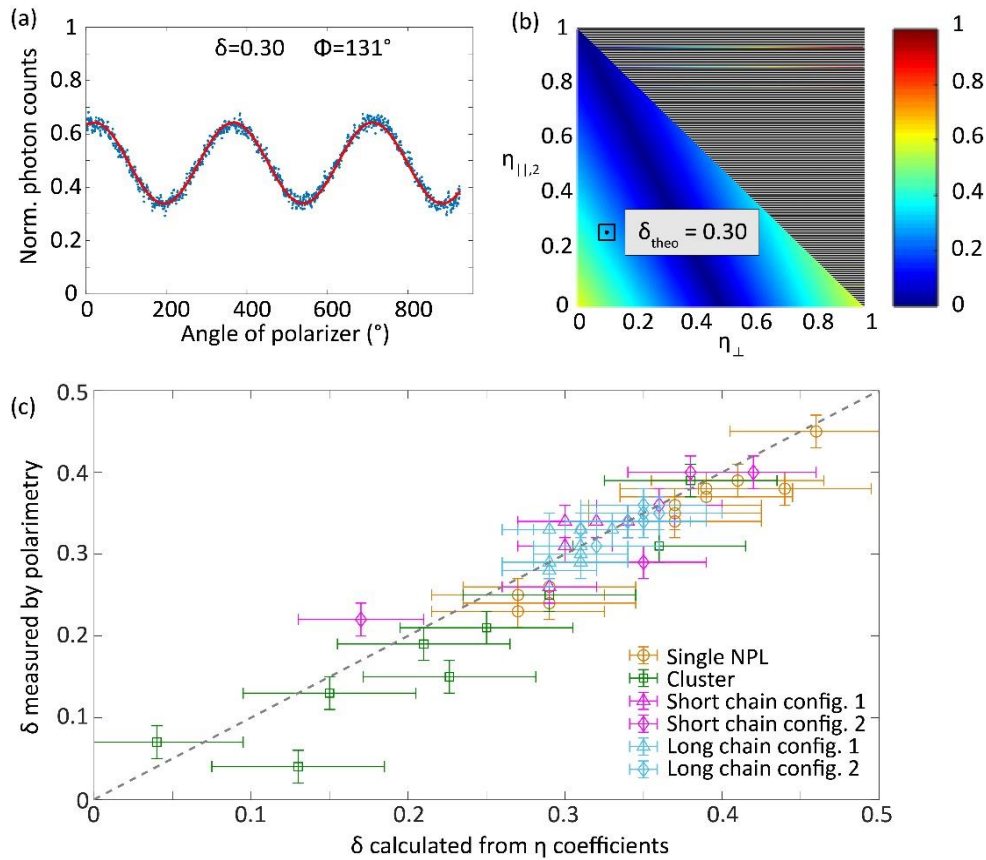

*Figure 3 : (a) Polarization analysis corresponding to the first chain in fig. 2(c). Blue dots: experimental data; red curve: cosine square function fitting. (b) Theoretical degree of polarization as a function of the components $\eta_{\parallel,2}$ and $\eta_\perp$. (c) Relation between the experimental degree of polarization $\delta$ and its theoretical value (calculated with the $\eta$ coefficients from Fourier imaging) for isolated NPLs, clusters*

*and chains (measured in two configurations : PMMA thicknesses 130 and 180 nm). The uncertainty for the experimental δ is ±0.02. The error bars for theoretical δ correspond resp. to ±0.05 and ±0.03 error on $\eta_{\parallel,2}$ or $\eta_\perp$.*

We considered as a reference a solution of dispersed (non assembled) NPLs deposited on a glass slide. Such CdSe platelets are known to deposit horizontally on the substrate[6]. For most emitters, the photon correlation function $g^{(2)}(\tau)$ (fig. 4(a), dotted line) measured in Hanbury-Brown and Twiss configuration showed clear antibunching (single photon emission). The ratio, here labelled $g_0$, between the area of the zero-delay peak and the average area of the other peaks, was in the range 0.06-0.40. This shows that multi-exciton radiative recombination is quenched by Auger effect, as $g_0$ is an estimate of the ratio between the biexciton and exciton quantum yield[38]. Given the area (around 140 nm²) of our platelets, this result is in agreement with ref. 39 : Ma et al. found $g_0$ close to 1 for NPLs larger than 150-200 nm² and decreased for smaller NPLs due to enhanced Auger interactions. For the emitter in fig. 4(a), we find $g_0 = 0.14$, indicating that we have a single platelet. The radiation pattern (fig. 4(b)) can be fitted by a sum of 2 horizontal (in-plane) dipoles, and the presence of a third vertical (out-of-plane) dipole can be excluded (see fig. S2(a) for precision analysis). Our measurement was further confirmed by comparing two measurement conditions, with or without a 30-nm PMMA film : we found $\eta_\perp = 0$ for the two configurations (fig. S2(b)).

Temporal post-selection can be used to remove the biexcitonic contribution (as explained in fig. S4). For some emitters however, we still find only incomplete antibunching ($g_0 > 0$) (fig. 4(c)), and in some cases no antibunching ($g_0 \sim 1$). We attribute these cases to clusters of platelets. It was seen already on fig. 1(a) that the isolated platelets tend to aggregate, parallel to each other. This is confirmed by decay and intensity-time trace curves (fig. S5) : a clear difference appears between the general emission properties of the single NPLs and the NPL cluster. The cluster radiation pattern in fig. 4(d) reveals two horizontal in-plane dipoles, showing that the clusters deposit horizontally like the single NPLs, but also a third vertical (out-of-plane) dipole which was not seen for the single NPLs. A similar out-of-plane component was found for all clusters.

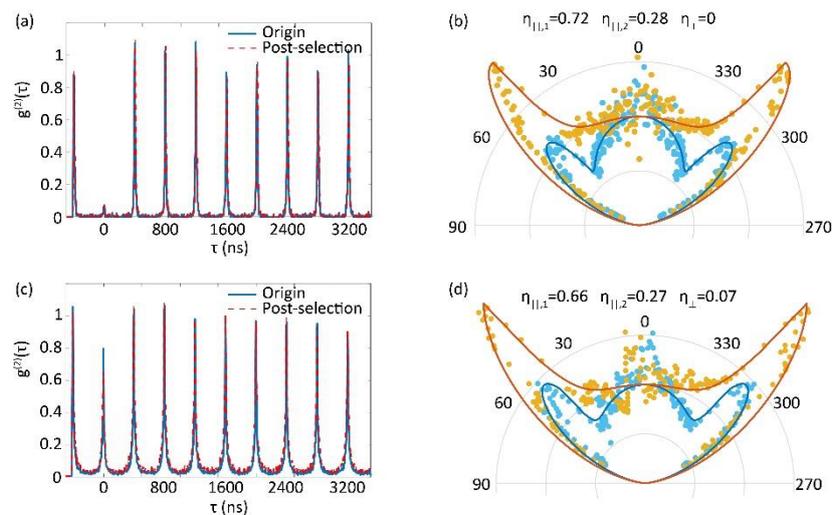

*Figure 4 : Photon correlation function $g^{(2)}(\tau)$ for (a) a single NPL and (c) a cluster. Full lines : original data, dotted lines : after post-selection treatment. (b) and (d) : Corresponding radiation patterns (dots) and theoretical fits (lines).*

Eventually, the measured dipoles coefficients $\eta$ are summarized in fig. 5 for single emitters of each type : long chains, shorter chains, clusters and single platelets. The main contributions are the two in-plane dipoles, parallel to the NPL plane, in agreement with our previous observations on various NPL geometries[6,35]. The in-plane dipolar anisotropy ratio $\eta_{\parallel,2}/\eta_{\parallel,1}$ is 0.43 for the single NPLs, 0.59 for the clusters and 0.42 for the chains : it seems quite similar for the different types of emitters and related mostly to the NPL rectangular shape. It is probably caused by the dielectric shape of the rectangles, which enhance the dipole along the NPL elongation axis[5-6,40]. For the clusters, the distribution of $\eta_{\parallel,2}/\eta_{\parallel,1}$ values is broader, possibly because, for the clusters, the platelets do not stack in well aligned assemblies, as can be seen on fig. 1(a) where the different rectangles are not parallel. The obtained values are similar for the two chain samples, showing no influence of either the chain length or its twist (TEM images show that the NPL rotation occurs in limited portions of the chains[26] and most NPLs form straight stacks, so the twisted chains still have a similar ratio $\eta_{\parallel,2}/\eta_{\parallel,1}$). The most striking difference between samples is that, while the NPL platelets show only in-plane dipoles (as was reported already by us[6,35] as well as other authors[3,5]), it is not the case for the NPL chains and clusters.

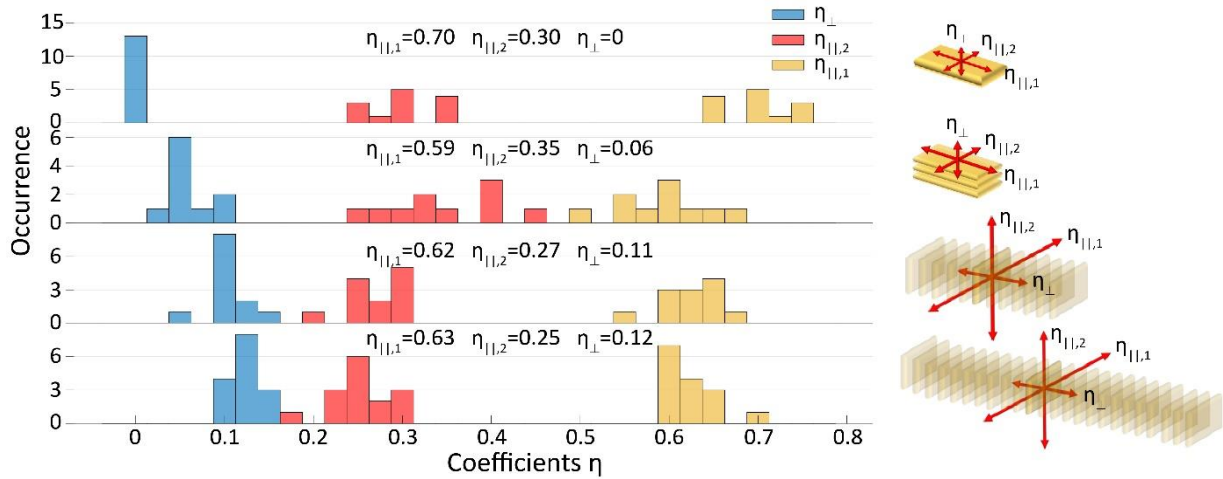

*Figure 5 : Histogram of the measured coefficients $\eta_{\parallel,1}$ (stronger in-plane dipole), $\eta_{\parallel,2}$ (lower in-plane dipole) and $\eta_\perp$ (out-of-plane dipole) for the different NPL samples : single NPLs, clusters, short chains and long chains. The average dipole contributions are given for each sample.*

Figure 6 plots the out-of-plane coefficient $\eta_\perp$ as a function of the number $N$ of emitters in the assembly. This number is of the order of 35 for the 100-500 nm chains and 250 for the 1-2 μm chains (note that not all 250 NPLs are excited by the laser spot). For the clusters, it can be estimated from the antibunching curves by assuming that they are a sum of $N$ identical single-photon sources so that $g_0 = 1 - 1/N$. The clusters with higher $g_0$ correspond to a large $N$ with large uncertainty and can be distinguished from the smaller clusters for which $g_0$ is significantly lower than 1.

Finally, figure 6 shows a continuous increase of the out-of-plane dipole with the number of NPLs in the stack : from $\eta \sim 0$ for the single platelets, to 0.05 for the small clusters ($N \sim 2-5$), to 0.1 for the larger clusters and 0.12 for the chains. We emphasize that it is the precision of our microphotoluminescence analysis protocol which allowed us to reveal this new dipole component, while it remained hidden in previous reports[4,17]. The origin of this out-of-plane component will be discussed below.

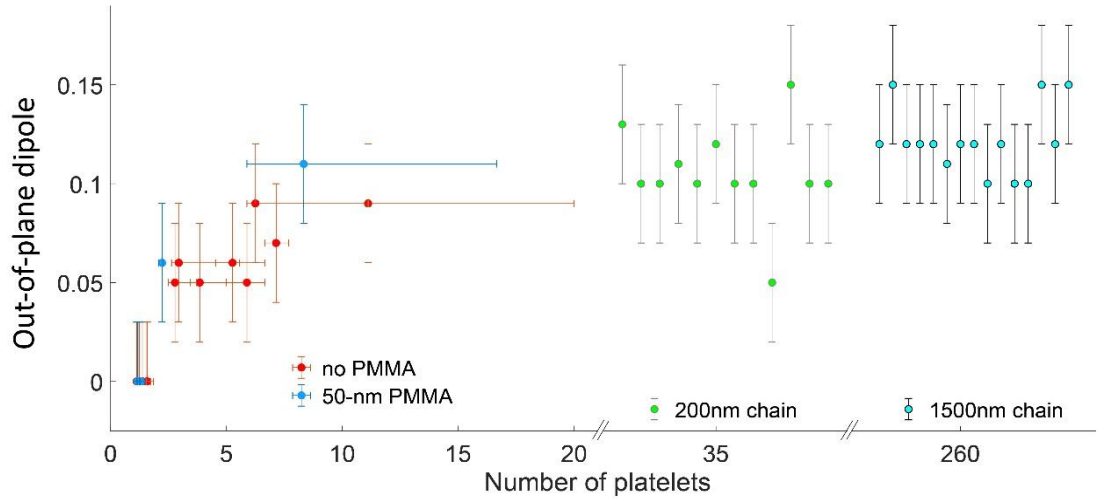

*Figure 6 : out-of-plane dipole component for various emitters, either isolated emitters (measured in the two configurations : with or without PMMA film) or chains, plotted as a function of the number of platelets in this emitter (obtained from $N = 1/(1 - g_0)$ for the isolated emitters – NPL or clusters – and from their length for the chains).*

**Discussion :**

A first hypothesis is that this dipole component originates from the elongated dielectric shape of the NPL chain out of a purely optical (nano-antenna) effect. Indeed for dipoles in an anisotropic dielectric structure the dipole component along the elongation axis is enhanced with respect to the other components[40]. This is probably the reason for the polarized emission of rectangular platelets[6]. We performed finite difference time domain (FDTD) simulations of the NPL chain (see full discussion in S.I. section C) and found that the dipole along the chain axis was indeed enhanced as the number of NPL increased. This effects begins to appear for just a few platelets and saturates around 5-10 platelets (fig. S6(b)). This corresponds to a length of 30-60 nm, when the NPL chain begins to be longer along the chain axis than along the axes of the NPL plane. However, because the chain is a succession of very thin high-index platelets and no just a µm-length block of high index, the antenna effect along the chain axis is not very strong. The enhancement achieved is only by 26% for the out-of-plane dipole, relative to the other dipoles. In order for the chain's out-of-plane dipole $\eta_\perp \sim 0.12$ to be explained only by a dielectric optical effect, the out-of-plane dipole of a single platelet would have to be $\eta_\perp \sim 0.08$ (see section C in S.I.), which is clearly excluded by our measurements. The emission of in-plane dipoles, even modified by the dielectric chain of the NPL chain, cannot account alone for our measurements (fig. S6(d)). Even though the FDTD model might not give a fully accurate description of the experimental chain structure, the order of magnitude of the dielectric optical effect is far from sufficient to explain our observations.

Having excluded optical effects, we turn to the electron-hole wavefunction. We first note that the excitation power is sufficiently low to ensure that only one exciton is excited in the whole chain per laser pulse (see S.I. – section D). The degree of polarization remained the same for all excitation powers, as well as the decay curve, while the emitted intensity was a linear function of the exictation power (figs. S7, S8, S9). Exciton-exciton interactions can thus be ruled out. We have also measured the decay curves and emission spectra along the two polarizations parallel and perpendicular to the chain and found no difference (fig. S10). We thus have no sign to support that the emerging out-of-plane dipole originates from a novel transition state.

The vertical confinement is extremely large for nanoplatelets so that the electron-light-hole transition is much higher in energy than the electron-heavy-hole transition[2]. Electron-heavy hole pairs can have ±1 or ±2 momentum[41]. The room-temperature emission originates only from the degenerate ±1 transitions[42] which is equivalent to an incoherent sum of two orthogonal in-plane dipoles. It is thus well understood why the electron-hole recombination in platelets shows no out-of-plane dipole. For nanocrystals, the exciton fine structure can theoretically present a third dipole component[43] but it is not the case for nanoplatelets.

Other emission mechanisms have been proposed in the literature, in particular in order to explain the second redshifted luminescence peak at low temperature[13] : phonon coupling[13], p-state emission[19], excimers[20], surface states[21], energy splitting by different dielectric confinements[44] and most recently trions[22]. Discussions of the different hypotheses and whether they are related to NPL stacking can be found in refs 20 and 22. In all experimental reports, this second peak contribution occurred only at low temperature (below 160 K)[21]. We have also observed this second peak on NPL chains at low temperatures (fig. S11). It clearly disappeared above 200 K so that this second peak is not contributing to our room-temperature measurements. We mention also that we have recently analysed FRET exciton diffusion within NPL chains at room temperature[24], but FRET is an incoherent non-radiative mechanism so that it should not modify the exciton wavefunction in a platelet.

It can also be noted that the platelets are not stacked exactly orthogonal to the chain axis : a slight tilt can be observed in the inset of fig. 1(b), so that a small component of the NPL dipole is in fact along the stacking axis. A statistical analysis of the TEM images shows that the NPL tilt angle is on average 7°, which, within a very simple geometric model, seems rather insufficient to induce the observed 0.12 out-of-plane dipole (see S.I. section G). However, a third dipole component may also appear if the NPL is not planar but slightly twisted. Such a twist would be difficult to observe on NPL-chain TEM images but has been demonstrated on single rectangular NPLs. It is caused by the internal strain due to non-uniform ligand coverage[26]. These strain and deformation are responsible for the helicoidal assembly of the NPL chains observed in fig. 1(b). Other authors have discussed the effect of NPL internal strain[45] and ligand coverage[46] on the emission wavelength. Recently, transient electric birefringence measurements[28] on chains of 20 to 90 NPLs have demonstrated a permanent dipole along the chain axis, which was attributed to such strain-induced shape deformation. It seems plausible that these deformations can also be responsible for our observed out-of-plane transition dipole.

**Conclusion:** We have analysed the luminescence of single nanoplatelets, clusters and chains of nanoplatelets. We used Fourier imaging to measure the different radiation dipole components. By comparing the result with polarization measurements and by comparing different measurement conditions (PMMA thicknesses), we were able to estimate the uncertainty to only ± 0.03 and 0.05 respectively for the out-of-plane (normal to the NPL plane) and minor in-plane contributions. For single platelets, we found no out-of-plane dipoles, in agreement with previous reports and with theoretical considerations. For NPL clusters and chains, on the other hand, we measured a minor but significant out-of-plane dipole, which increased generally as a function of the number of platelets. We discussed various mechanisms for the presence of this dipole and suggest that it might be related to a strain-induced NPL deformation within the NPL assembly, with an enhancement by a dielectric antenna effect. A better understanding of the role of self-assembly on the emission dipole could be obtained by further high-resolution electron imaging and a comparison with numerical models of the NPL crystalline structure and internal strain. We emphasize that it is our dipole analysis protocol, combining Fourier imaging and polarimetry, which allows to probe dipole components in different emitting systems with record precision thanks to the design of optimal experimental configuration (such as PMMA thickness). The method which is demonstrated here can have a wide range of applications in fundamental study of optical and electrical properties of semiconducting single emitters or two-dimensional materials, such as perovskite quantum dots and transition metal dichalcogenide semiconductors.

**Methods :**

Sample preparation :

*Nanoplatelets synthesis*: first we prepared cadmium oleate from a mixture of dissolved Sodium oleate and Cadmium nitrate tetrahydrate through a careful washing and drying process. Then we introduced cadmium oleate with commercial cadmium oleate and ODE in a three-neck flask equipped with a septum, a temperature controller and an air condenser. We carefully controlled the temperature and reaction ambience and added cadmium acetate dihydrate and oleic acid successively. After cooling the flask, we obtained a mixture containing 5-ML CdSe NPL, 3ML-CdSe NPL and quantum dots in solution, from which we separated 5-ML CdSe NPL by centrifugation.

*Self-assembly of NPLs chains*: an appropriate amount of 5ML-CdSe NPL solution was diluted in hexane and Oleic acid was then added, the amount of which is crutial to the length of assembled NPLs chains. The sample was sonicated for 10 minutes and the solvent was slowly evaporated. At this point we obtained highly-ordered NPLs chains.

More details about the synthesis and assembly protocol can be found in ref. 24.

Fluorescence microscopy : we performed optical measurements using a homebuilt inverted fluorescence microscope equipped with a laser scanning system. To image the NPLs chain, we used a mercury lamp to achieve a wide field excitation and used a CCD camera (QImaging Retiga EXi, pixel size 6.45 µm correspondin to 72nm on the image) for detection. A same objective (Olympus apochromat 100X 1.4N.A.), mounted on a piezo-electric stage, was used to focus the 470-nm diode laser beam (PDL 800-D PicoQuant, 70-ps pulses, 2.5-MHz rate) on the sample plane to scan/excite a single emitter and also to collect its emission. The scattered excitation light was filtered by a set of filters and only the 549 nm fluorescence light can reach the detectors.

During all measurements, the laser excitation power ranged typically from 5- to 7.5-nW (resp. 3- to 6-nW) for single NPLs (resp. assembled NPLs chains), within the linear excitation regime of the emitters.

To prepare the samples for optical study, the solution of single NPLs (respectively NPLs chains) was diluted 10000 times (resp. 500 times) in hexane and spin-coated on a glass slide at 4000 rpm for 40 s to get well-seperated single emitters. The deposited sample was then covered by a layer of PMMA with different thickness ranging from 130 to 210 nm as determined by profilometry. For theoretical calculations, the PMMA, glass and immersion oil will all be taken equal to 1.5.

Fourier plane imaging : to record the Fourier plane image, we used a Fourier lens to conjugate the rear focal plane (Fourier plane) of the objective to a EMCCD (Andor iXon Ultra 897) as described in details in ref. 35. The magnification of the conjugated image on the camera is 1.05X, with angular resolution ranging between 0.3° and 0.8° (corresponding to a collection angle from 0° to 69°).

Polarization measurement : we put a rotating half wave plate in front of a polarizing beam splitter cube (50:50), working equivalently to a rotating polarizer with angle denoted as α. The emission light from the microscope passed through this rotating system, was separated into reflection/transmision paths and detected by 2 avalanche photodiodes (PerkinElmer SPCM), respectively. We plot the normalized intensity on one photodiode as:

$$I_{exp}(\alpha) = \frac{I_1}{I_1 + I_2}$$

in which $I_1$ and $I_2$ are the measured intensities on the two photodiodes. We use a $cos^2(\alpha)$ function to fit $I_{exp}(\alpha)$ as:

$$I_{fit}(\alpha) = k_1 + k_2 \cdot cos^2(\Phi - \alpha)$$

from which we can extract the polarization orientation $\Phi$ and calculate the experimental degree of polarization $\delta$ by:

$$\delta = \frac{I_{max} - I_{min}}{I_{max} + I_{min}} = \frac{k_2}{k_1 + 2k_2}$$

Theoretical polarization calculations : if a polarizer is rotated along orientation $\alpha$ in front of the detection photodiode, each of our three considered dipoles has a detected intensity :

$$I_{\parallel,1}(\alpha) = A + C \cdot cos^2(\Phi - \alpha)$$
$$I_{\parallel,2}(\alpha) = A + C \cdot cos^2(\Phi + \frac{\pi}{2} - \alpha)$$
$$I_{\perp}(\alpha) = B$$

where $A, B$ and $C$ are the coefficients defined in ref. 36. The theoretical degree of polarization for an incoherent sum of the three dipoles is thus :

$$\delta = \frac{I_{max} - I_{min}}{I_{max} + I_{min}}$$

in which

$$I_{max} = \eta_{\parallel,1} \cdot (A + B) + \eta_{\parallel,2} \cdot A + B$$
$$I_{min} = \eta_{\parallel,1} \cdot A + \eta_{\parallel,2} \cdot (A + C) + B$$

eventually we get:

$$\delta = \frac{C \cdot (\eta_{\parallel,1} - \eta_{\parallel,2})}{2A + C + \eta_{\perp} \cdot (2B - 2A - C)}$$

Photon correlation function measurement:
We measured the photon correlation function of isolated emitters under Hanbury-Brown and Twiss configuration. We used the 470-nm pulsed laser (70ps pulse width) with 2.5MHz repitation rate to excite a single emitter and used the polarizing beam splitter cube (50:50) to separate the emitted photon flux into reflection/transmision paths to be detected by 2 avalanche photodiodes. The second-order cross correlation function $g^{(2)}(\tau)$ is defined as

$$g^{(2)}(\tau) = \frac{\langle I_1(t)I_2(t+\tau)\rangle}{\langle I_1(t)\rangle\langle I_2(t+\tau)\rangle}$$

where $I_1(t)$ and $I_2(t)$ are the signal intensities on the two photodiodes and $\tau$ is the delay time.